\documentclass[aps,pra, onecolumn,superscriptaddress,10pt,floatfix,amsmath,amssymb]{revtex4-2}

\usepackage{graphicx} 
\usepackage{dcolumn}
\usepackage{xcolor}
\usepackage{bm}
\usepackage{physics}
\usepackage{lipsum}  
\usepackage{hyperref}
\usepackage[normalem]{ulem}
\graphicspath{{images/}}

\newcommand{\sect}[1]{\textit{#1}.-- }

\begin{document}

\title{Thermodynamics and protection of discrete time crystals}

\date{\today}
    
\author{Gabriele Cenedese}
    \email{gcenedese@uninsubria.it}
    \affiliation{Center for Nonlinear and Complex Systems, Dipartimento di Scienza e Alta Tecnologia, Universit\`a degli Studi dell'Insubria, via Valleggio 11, 22100 Como, Italy} 
    \affiliation{Istituto Nazionale di Fisica Nucleare, Sezione di Milano, via Celoria 16, 20133 Milano, Italy}

\author{Samuel T. Mister}
    \affiliation{Quantum Engineering Technology Labs and Quantum Engineering Centre for Doctoral Training,
H. H. Wills Physics Laboratory and Department of Electrical and Electronic Engineering, University of Bristol, Bristol BS8 1FD, United Kingdom} 

\author{Mauro Antezza}
    \affiliation{Laboratoire Charles Coulomb (L2C) UMR 5221 CNRS-Universit\'e de Montpellier, F- 34095 Montpellier, France}
    \affiliation{Institut Universitaire de France, 1 rue Descartes, F-75231 Paris Cedex 05, France}
\author{Giuliano Benenti}
    \affiliation{Center for Nonlinear and Complex Systems, Dipartimento di Scienza e Alta Tecnologia, Universit\`a degli Studi dell'Insubria, via Valleggio 11, 22100 Como, Italy} 
    \affiliation{Istituto Nazionale di Fisica Nucleare, Sezione di Milano, via Celoria 16, 20133 Milano, Italy}
\author{Gabriele De Chiara}
    \affiliation{Física Teòrica: Informació i Fenòmens Quàntics, Departament de Física, Universitat Autònoma de Barcelona, 08193 Bellaterra, Spain}
 \affiliation{Centre for Quantum Materials and Technology, School of Mathematics and Physics, Queen’s University Belfast, Belfast BT7 1NN, United Kingdom}
   
\begin{abstract}
Discrete-Time Crystals (DTC) are a non-equilibrium phase of matter characterized by the breaking of time-translation symmetry in periodically driven quantum systems. In this work, we present a detailed thermodynamic analysis of DTCs in a one-dimensional spin-1/2 chain coupled to a thermal bath. We derive a master equation from the microscopic model, and we explore key thermodynamic quantities, such as work, heat, and entropy production. Our results reveal that the DTC signature inevitably decays in the presence of environmental noise, but we show that a periodic measurement scheme can mitigate the effects of decoherence, stabilizing the subharmonic oscillations of the DTC for extended periods. These findings provide insights into the robustness of time-crystalline phases and potential strategies for protecting them in experimental settings.
\end{abstract}
\maketitle
\section{Introduction}
The study of out-of-equilibrium systems has received a lot of attention in recent years thanks to the advancements in the control and manipulation of experimental quantum platforms for the simulation of many-body systems \cite{daley2014quantum,georgescu2014quantum,daley2022practical,eisert2015quantum}. In this context, quantum time crystals \cite{wilczek2012quantum}, exhibiting time translation symmetry breaking, are at the centre of many recent works, thanks to their fascinating role in the statistical mechanics of non equilibrium phenomena as well as their applications as time sensors \cite{PhysRevA.109.L050203, PhysRevLett.132.050801, montenegro2023quantum, zheludev2024time}. Time crystals are usually classified as continuous or discrete time crystals, depending on the breaking, continuous or discrete, of the time translation symmetry.  
Discrete-time crystals (DTC) \cite{sacha2017time,else2020discrete,zaletel2023colloquium}, which will be the main subject of this paper, are periodically driven systems, usually involving disordered localized many-body systems, which respond with an order parameter with periodicity which is an integer multiple of the driving periodicity. Discrete-time crystals have been recently realised experimentally in systems of NV centres in diamond \cite{choi2017observation, beatrez2023critical}, trapped ions \cite{zhang2017observation}, optical systems \cite{taheri2022all} and superconducting systems \cite{ippoliti2021}. Their continuous versions has been realised in cold and ultracold systems \cite{kongkhambut2022observation,autti2018observation, ferioli2023non}. 

Despite the huge interest that time crystals have generated, the analysis of their thermodynamics, for instance how much work is necessary to maintain them, their heat exchange and entropy production, has been mostly overlooked. In Ref.\cite{carollo2024quantum}, the thermodynamics of the so called boundary time crystals, a class of continuous dissipative time crystals \cite{iemini2018boundary}, has been thoroughly studied by introducing a microscopic model and a consistent framework for the thermodynamic assessment of their dynamics. In Ref.\cite{paulino2024thermodynamics}, the analysis was extended to two coupled boundary time crystals and it was shown how the system can be used for energy storage.

In this paper, we present an extensive thermodynamic analysis of DTC. We introduce the two-stroke spin-1/2 spin chain Hamiltonian and a thermodynamically consistent Lindblad master equation for the description of its dissipative evolution. Using this framework, in the first part of the paper, we evaluate the work and heat exchanged during each period of the time evolution as well as the entropy production. In the second part, we introduce a periodic measurement scheme with the objective of preserving as much as possible the time crystal signature that would otherwise disappear in the presence of an external environment. We show that it is possible to significantly mitigate the damaging effects of the environment on the time crystal, thus preserving the subharmonic oscillations characteristic of DTCs. Although the benefits of the measurement scheme are limited on average, we observe a stabilisation of parameters associated to the staggered magnetisation at the level of individual trajectories. 

The paper is organized as follows: In section \ref{Sec:1} we present the DTC model that is the subject of our studies and we provide a pedagogical derivation from the microscopic model of the system master equation under the assumption of weak coupling and Markovian evolution. In sections \ref{Sec:3} and \ref{Sec:4} we show our results concerning the subharmonic response of the DTC in the presence of a thermal bath and its thermodynamic properties, in particular we show that the signature of the DTC will perish after a certain time regardless of temperature and coupling. In section \ref{Sec:5} we describe the measurement scheme and show our results. Finally, in section \ref{Sec:6} we draw our conclusions.

\section{Model}
\label{Sec:1}
\sect{Time-Crystal model}
In this paper we study a driven one-dimensional spin-1/2 chain described by the time-dependent Hamiltonian \cite{else2016floquet,khemani2016phase, lazarides2017fate, yao2017discrete}:
\begin{equation}
    H_S(t)=
    \begin{cases}
    H_z \,\,\,\,\text{if}\,\,\,\, 0\leq t <T_z,  \\
    H_x \,\,\,\,\text{if}\,\,\,\, T_z \leq t < T=T_x +T_z.
    \end{cases}
\end{equation}
This Hamiltonian is a periodic function with period $T=T_z+T_x$, where $H_z$ and $H_x$ are defined as follows: 
\begin{eqnarray}
    H_z&=&\frac{1}{2}\sum_{i=1}^{N} h_i \sigma_i^z,  
\\
    H_x&=&\sum_{i=1}^{N-1} J_i \sigma_i^x \sigma_{i+1}^x, 
\end{eqnarray}
where $(\sigma_i^x,\sigma_i^y,\sigma_i^z)$ represent the Pauli operators for spin $i$.
The values of the magnetic fields $h_i$ and couplings $J_i$ are randomly drawn from uniform distributions with mean values $\bar{J}$ and $\bar{h}$:
\begin{eqnarray}
J_i&=&\bar{J} + \delta J \xi_i,
\\
h_i&=&\bar{h} +\delta h \tilde{\xi}_i,
\end{eqnarray}
where $\xi_i$ and $\tilde{\xi}_i$ are independent uniformly distributed random variables between -1 and 1, $\delta h$ and $\delta J$ are the widths of the distributions. From now on we will refer to the part of the evolution defined by $H_x$ ($H_z$) as the $x$-stroke ($z$-stroke).
To observe the DTC effects we need to consider this model in the $\pi$-SG phase \cite{khemani2016phase} (SG=spin glass). Additionally the initial state of the system must break the $\mathbb{Z}_2$ symmetry of the Hamiltonian. 
We have numerically implemented this model for an $N$-spin system.
Henceforth, we will fix the mean value of $h_i$ as $\bar{h}=\pi$ and of $J_i$ as $\Bar{J}=\pi/4$, the width of the distribution is fixed as $0.5$ for both the set of parameters. 
The inital state is fixed as $\rho(0)=\ket{+}\bra{+}$, where $\ket{+}$ being the state of all the spins pointing upwards along the $x$ direction: $\ket{+}=\ket{\uparrow_x\dots\uparrow_x}$. By choosing these parameters and initial state and by setting $ T_z = T_x = 1 $, the dynamics of the time crystal becomes readily comprehensible: the $ z $-stroke simply rotates each spin by $ \theta_i = h_i T_z $ around the $ z $-axis, then the system evolves under $ H_x $. Here, disorder is needed to keep the system in a many-body localized phase; otherwise, the system would thermalize, destroying the time crystal ordering. The subsequent $ z $-stroke approximately returns the spins to their initial positions. Finally, the $ x $-stroke localizes the spins again.
The emergence of the time crystal can be explored by measuring either the magnetization of a single spin $\langle \sigma_i^x(t) \rangle$ or the total magnetization $\langle S_x(t) \rangle = \frac{1}{N}  \sum_i \langle\sigma_i^x(t) \rangle$, both quantities exhibiting periodicity at half the frequency of the driving.
This specific choice of the initial state provides a nice way to visualise these dynamics but we must note that this periodic behaviour is observed for many initial states, however, the amplitude of the signal will depend on the initial value of the chosen signature.


\sect{Derivation of the master equation} With this setup in mind, we can investigate the role of an external environment on the time crystal behaviour, following the standard approach of open quantum systems weakly coupled to their environment (see for instance Ref.~\cite{breuer2002theory}).
In the presence of an external bath, the total Hamiltonian of the system transforms into:
\begin{equation}
    H(t)=H_S(t)+ H_{B} + H_I,
\end{equation}
where $ H_S (t) $ represents the time-dependent time-crystal Hamiltonian, $H_B$ denotes the bath Hamiltonian, and $H_I$ the interaction term. Since $H_S(t)$ consists of two temporally constant alternating terms, we can effectively treat it as constant during each stroke. This allows us to analyze the two strokes separately and for this reason we will henceforth consider $H_S$ as time-independent. 
Starting from the interaction picture von Neumann equation, and assuming without loss of generality $\text{Tr}_B \left[H_I(t), \rho(0) \right]=0$, the quantum master equation for the system density matrix reads:
\begin{equation}
    \label{eq:initialme}
    \frac{\text{d}}{\text{d}t}\rho_S(t)=-\int_0^t \text{d}s \, \text{Tr}_B \left[ H_I(t),\left[ H_I(s),\rho(s)\right]\right],
\end{equation}
where $\rho$ represents the joint system-bath density matrix, and $\rho_S$ denotes the system density matrix obtained by tracing out the reservoir degrees of freedom, by employing the partial trace $\text{Tr}_B$, and finally $H_I(t)$ is the time-dependent interaction picture version of $H_I$ (from now on, we will refer to operators without explicit time dependence as operators in the Schrödinger picture, and those with explicit time dependence as operators in the interaction picture). Here we have assumed units in which $\hbar=1$. We assume that the coupling between the system and the reservoir is weak, allowing us to apply the Born approximation. In this case, the system-bath density matrix can be approximated as separable: $\rho(s) \simeq \rho_S(s) \otimes \rho_B$, where $\rho_B$ is the initial state of the reservoir assumed constant (this condition ensures that implementing the evolutions of individual strokes separately, and therefore neglecting the correlations generated between the environment and the system in previous strokes does not introduce any errors). To further simplify, we perform the Markov approximation by replacing the integrand $\rho_S(s)$ with $\rho_S(t)$ in Eq.~\eqref{eq:initialme}. This ensures that the evolution of the system density matrix depends only on the state $\rho_S(t)$ at time $t$. As a result, we obtain the so-called Redfield equation, which is local in time but not yet fully Markovian. To derive the Markovian master equation, we replace $s$ with $t-s$ in the integrand and extend the upper limit of the integral to infinity. This approximation is valid when the timescale $\tau_R$ over which the open system's state varies is much larger than the reservoir's characteristic correlation time $\tau_B$. At this point the master equation reads:
\begin{equation}
    \frac{\text{d}}{\text{d}t}\rho_S(t)=-\int_0^\infty \text{d}s \, \text{Tr}_B \left[ H_I(t),\left[ H_I(t-s),\rho_S(t)\otimes\rho_B\right]\right].
\end{equation}
At this stage, we can perform the rotating wave approximation (RWA) by averaging over the rapidly oscillating terms. To achieve this, we have to explicitly introduce the interaction Hamiltonian, which, without loss of generality, can be written, in the Schr\"odinger's picture, as:
\begin{equation}
    H_I=\sum_\alpha A_\alpha\otimes B_\alpha,
\end{equation}
where $A_\alpha$ and $B_\alpha$ are two sets of Hermitian operators of the system and the bath, respectively. 
Let us define the eigenstates of the system's Hamiltonian: $H_S\ket\epsilon=\epsilon\ket\epsilon$, with energy eigenvalues $\epsilon$.
The RWA is easily performed by decomposing the interaction Hamiltonian $H_I$ into the energy eigenbasis $\{\ket\epsilon\}$:
\begin{equation}
    A_\alpha(\omega)\equiv \sum_{\epsilon'-\epsilon=\omega}  \ket{\epsilon}\bra{\epsilon} A_\alpha \ket{\epsilon'}\bra{\epsilon'},
\end{equation}
where the sum is restricted to all energies $\epsilon$ and $\epsilon'$ with a fixed difference $\omega$. We notice that $\sum_\omega A_\alpha(\omega)=A_\alpha$ by using the completeness relation and that these operators are no longer Hermitian: $A_\alpha^\dagger(\omega)=A_\alpha(-\omega)$. It is easy to see that the interaction picture version of $A_\alpha(\omega)$ can be written as
\begin{equation}
    A_{\alpha,\omega}(t)=e^{iH_St} A_\alpha(\omega) e^{-i H_S t}=e^{-i\omega t} A_\alpha(\omega),
\end{equation}
which allows us to rewrite the interaction Hamiltonian, in the interaction picture, as
\begin{equation}
    H_I(t)=\sum_{\alpha,\omega} e^{-i\omega t} A_\alpha (\omega) \otimes B_\alpha(t)=\sum_{\alpha,\omega} e^{i\omega t} A_\alpha^\dagger (\omega) \otimes B^\dagger_\alpha(t),
\end{equation}
where $B_\alpha (t) = e^{iH_B t} B_\alpha e^{-iH_B t}$ is the bath interaction term in the interaction picture. By inserting $H_I(t)$ into the master equation we obtain after some algebra:
\begin{equation}
    \frac{\text{d}}{\text{d}t}\rho_S(t)= \sum_{\omega \omega'} \sum_{\alpha \beta} e^{i(\omega'-\omega)t}\Gamma_{\alpha \beta}(\omega)\left[ A_\beta(\omega)\rho_S(t) A_\alpha^\dagger(\omega')-A^\dagger_\alpha(\omega')A_\beta(\omega)\rho_S(t)\right] + \text{h.c.},
\end{equation}
where h.c. stands for the Hermitian conjugated, $\Gamma_{\alpha \beta }(\omega)\equiv \int_0^\infty \text{d}s e^{i\omega s} \langle B_\alpha^\dagger (t)B_\beta(t-s)\rangle$ is the Fourier transform of the reservoir correlation functions:
\begin{equation}
    \langle B_\alpha^\dagger (t)B_\beta(t-s)\rangle=\langle B_\alpha^\dagger (s)B_\beta(0)\rangle\equiv\text{Tr}\left[ B_\alpha^\dagger (s) B_\beta(0)\rho_B \right],
\end{equation}
where the first equality holds assuming $\rho_B$ as a stationary state of the reservoir. The RWA consists in neglecting rapidly oscillating terms for which $\omega\neq\omega'$. The typical timescale $\tau_S$ of the system's intrinsic evolution is given by the characteristic value of $|\omega - \omega'|^{-1}$. If $\tau_S$ is small compared to the relaxation time $\tau_R$, the terms $e^{i(\omega'-\omega)t}$ oscillates rapidly compared to the timescale over which the system changes appreciably. Therefore, these terms can be neglected, resulting in the following master equation:
\begin{equation}
    \frac{\text{d}}{\text{d}t}\rho_S(t)= \sum_{\omega} \sum_{\alpha \beta} \Gamma_{\alpha \beta}(\omega)\left[A_\beta(\omega)\rho_S(t) A_\alpha^\dagger(\omega)-A^\dagger_\alpha(\omega)A_\beta(\omega)\rho_S(t)\right] + \text{h.c.}.
\end{equation}
In general we can write $\Gamma_{\alpha \beta}(\omega)=\frac{1}{2}\gamma_{\alpha \beta}(\omega)+ i S_{\alpha \beta}(\omega)$, where, for fixed $\omega$, $\gamma_{\alpha \beta}(\omega)= \int_{-\infty}^\infty \text{d}s e^{i\omega s} \langle B_\alpha^\dagger (s)B_\beta(0)\rangle$ is a positive matrix and $S_{\alpha \beta}(\omega)$ is a Hermitian matrix. With these definitions the master equation becomes:
\begin{equation}
    \frac{\text{d}}{\text{d}t} \rho_S(t) = -i \left[ H_{LS},\rho_S(t) \right]  + \sum_{\omega} \sum_{\alpha \beta} \gamma_{\alpha \beta}(\omega)   \biggl( A_\beta(\omega) \rho_S(t) A_\alpha^\dagger(\omega)-  \frac{1}{2} \left\{ A^\dagger_\alpha(\omega)A_\beta(\omega),\rho_S(t)\right\} \biggr),
    \label{eq:ME}
\end{equation}
where the first term gives a Hamiltonian contribution to the dynamics and contains the so-called Lamb shift Hamiltonian $H_{LS}=\sum_\omega \sum_{\alpha \beta}S_{\alpha \beta}(\omega)A^\dagger_{\alpha}(\omega) A_\beta(\omega)$, the second term describes the dissipation and it is usually known as the dissipator. The Schrödinger picture master equation is simply obtained by adding the system Hamiltonian to the unitary part of Eq.~\ref{eq:ME}. In this way $H_{LS}$, since it commutes with $H_S$, effectively acts as a renormalization of the energy levels of $H_S$ due to the coupling between the system and the reservoir. However, in many cases, the contribution of $H_{LS}$ is negligible, for this reason, from this point forward, we will neglect $H_{LS}$ for the sake of simplicity.

Eq. \ref{eq:ME} can be brought to the so-called Lindblad form by diagonalizing $\gamma_{\alpha \beta}(\omega)$:
\begin{equation}
    \frac{\text{d}}{\text{d}t} \rho_S(t) = -i \left[ H_{S},\rho_S(t) \right]  + \sum_{\omega} \sum_{\alpha} \Tilde{\gamma}_{\alpha}(\omega)   \biggl( \Tilde{A}_\alpha(\omega) \rho_S(t) \Tilde{A}_\alpha^\dagger(\omega)-  \frac{1}{2} \left\{ \Tilde{A}^\dagger_\alpha(\omega)\Tilde{A}_\alpha(\omega),\rho_S(t)\right\} \biggr),
    \label{eq:LME}
\end{equation}
where $\Tilde{A}_\alpha(\omega)$ are related to $A_\alpha(\omega)$ by the diagonalization matrix.
Up to this point the derivation of the master equation is general; now we explicitly model the bath to get an expression for $\gamma_{\alpha \beta}(\omega)$. We couple the time-crystal to a bath of harmonic oscillators:
\begin{equation}
    H_{B} =\sum_{k=1}^{\infty}\omega_k b_k^\dagger b_k,
\end{equation}
where the operators $b_k$ ($b_k^\dagger$) represent the annihilation (creation) operators of oscillator $k$, $\omega_k$ being the charateristic frequency of the k-th mode. The bath initial state is the Gibbs thermal state at inverse temperature $\beta$: $\rho_B=\text{exp}(-\beta H_B)/\mathcal{Z}$, where $\mathcal{Z}=\text{Tr}[\text{exp}(-\beta H_B)]$ is the partition function.
As an interaction Hamiltonian we consider the following, which is fairly standard in the spin-bosons model case (see \cite{breuer2002theory,rivas2012open,albash2012}), where we assume that the oscillators couple to each spin in the same way:
\begin{equation}
    H_{I} = \sum_{i=1}^{N} \sigma^\alpha_i \otimes  \sum_{k=1}^\infty g_k (b_k^\dagger+ b_k)  \equiv V \otimes B,
\end{equation}
where $g_k$ is the bath coupling strength between the $k$-th mode and the time-crystal. Here, we have not yet specified which spin direction is coupled to the bath and the index $\alpha$ is left generic for this reason. Under these assumptions it is straightforward to show (see Appendix H of \cite{albash2012} for the explicit calculation) that the reservoir correlation function takes the following form:
\begin{equation}
    \langle B(t)B(0)\rangle= \sum_{k=0}^\infty \frac{g_k^2}{1-e^{-\beta \omega_k}} \left( e^{-i\omega_k t } + e^{i\omega_k t-  \beta \omega_k}\right)= \int_{0}^\infty \text{d}\Tilde{\omega}\, \frac{J(\Tilde{\omega})}{1-e^{-\beta \Tilde{\omega}}} \left( e^{-i\Tilde{\omega}t } + e^{i\Tilde{\omega} t-  \beta \Tilde{\omega}}\right),
\end{equation}
where, after the second equality, we have replaced the sum over $k$ with an integral over $\Tilde{\omega}$ by defining the bath spectral function $J(\Tilde{\omega})\equiv\sum_k g_k^2 \, \delta(\Tilde{\omega}-\omega_k)$, which characterizes how different frequencies of the bath modes couple to the system. We want to stress that in this expression $\Tilde{\omega}$ refers to the frequencies of the environment modes, and thus $\Tilde{\omega}\in [0, +\infty)$.
At this point, we only need to calculate the Fourier transform of this expression to derive the rates $\gamma(\omega)$:

\begin{eqnarray}
\gamma(\omega)&=&\int_{-\infty}^{\infty} \text{d}t\, e^{i\omega t} \int_{0}^\infty \text{d}\Tilde{\omega}\, \frac{J(\Tilde{\omega})}{1-e^{-\beta\Tilde{\omega}}} \left( e^{-i\Tilde{\omega} t } + e^{i\Tilde{\omega} t-  \beta \Tilde{\omega}}\right)=
\\
&=&\int_{0}^\infty \text{d}\Tilde{\omega}\, 2\pi J(\Tilde{\omega}) \biggl[  \left[ n(\Tilde{\omega})+1\right] \delta(\omega-\Tilde{\omega})+ n(\Tilde{\omega})\delta(\omega+  \Tilde{\omega}) \biggr],
\end{eqnarray}
where, from the first to the second line, we calculated the Fourier transform using the properties of the Dirac's delta function and we defined $n(\tilde{\omega}) = (e^{\beta \tilde{\omega}} - 1)^{-1}$, that is the thermal occupation number of a bosonic mode with energy $\tilde{\omega}$. This last equation offers an interesting physical interpretation: one finds two terms, both proportional to the spectral density, the first term describes emission of energy into the environment while the second describes the probability of absorption of energy from the environment, which is only possible if a mode of energy $\omega$ is occupied (absorption occurs when $\omega<0$). Furthermore, this final integral can be computed exactly, yielding:
\begin{eqnarray}
    \gamma(\omega)&=&2\pi\biggl[ J(\omega) \left[ n(\omega)+1 \right] \Theta(\omega) + J(-\omega) n(-\omega) \Theta(-\omega) \biggr] = 
    \\
    &=& 2\pi J(|\omega|) \biggl[  \left[ n(|\omega|)+1 \right] \Theta(\omega) +  n(|\omega|) \Theta(-\omega) \biggr]  =
    \\
    &=&\frac{2\pi J(|\omega|)}{1-e^{-\beta |\omega|}}(\Theta(\omega)+e^{-\beta|\omega|}\Theta(-\omega)),
\end{eqnarray}
where $\Theta(\omega)$ is the Heaviside function.
Consequently, we now explicitly model the system-bath coupling using an Ohmic spectral function $J(\omega) = \eta |\omega| e^{-|\omega|/\omega_c}$, characterized by the Kondo parameter $\eta$, that quantifies the system-bath coupling strength, and the cut-off frequency $\omega_c$. For simplicity and without loss of generality, we define $2\pi \eta\equiv \Gamma^2$ and we fix $\omega_c=1$.  Hence, the free parameters of our model are the coupling strength $\Gamma$ and the inverse temperature $\beta$.

To summarize, the master equation can be written as:
\begin{equation}
    \frac{\text{d}}{\text{d}t} \rho_S(t) = -i \left[ H_{S},\rho_S(t) \right]  + \sum_{\omega}   \biggl( L_\omega \rho_S(t) L_\omega^\dagger-  \frac{1}{2} \left\{ L_\omega^\dagger L_\omega,\rho_S(t)\right\} \biggr),
    \label{eq:LMEfinal}
\end{equation}
with the jump operators $L_\omega$ defined as:
\begin{equation}
    L_\omega=\sqrt{\gamma(\omega)} \sum_{\epsilon'-\epsilon=\omega}  \ket{\epsilon}\bra{\epsilon} V \ket{\epsilon'}\bra{\epsilon'}= \sqrt{\gamma(\omega)} \sum_{\epsilon'-\epsilon=\omega}  |V_{\epsilon\epsilon'}|\ket{\epsilon}\bra{\epsilon'},
\end{equation}
with $\omega=\epsilon'-\epsilon$, $\ket{\epsilon}$ ($\ket{\epsilon'}$) being an eigenstate of $H_S$ (which can be either $H_z$ or $H_x$) with energy $\epsilon$ ($\epsilon'$) and finally $V_{\epsilon\epsilon'}=\bra{\epsilon}V\ket{\epsilon'}$. 

Since it will be useful later, let us rewrite the master equation for a single stroke as $\dot{\rho_S} = \mathcal{L}_{x(z)} \rho_S$, where $\mathcal{L}_{x(z)}$ is the Lindblad super-operator of the system in the $H_x (H_z)$ stroke. In this framework, the evolution of the system's density matrix is given by the exponential of $\mathcal{L}_{x(z)}$, introducing the evolution super-operator for each strokes as $\Lambda_{x(z)}(t) \equiv e^{t \mathcal{L}_{x(z)}}$. For simplicity, we define the evolution super-operator for a single period $T$ as:
\begin{equation}
    \Lambda(t)\equiv
    \begin{cases}
    \Lambda_z(t) \,\,\,\,\text{if}\,\,\,\, 0\leq t <T_z,  \\
    \Lambda_x(t-T_z)\Lambda_z(T_z) \,\,\,\,\text{if}\,\,\,\, T_z \leq t \leq  T,
    \end{cases}
\end{equation}
where $0\leq t \leq T$.

\section{Results} 
\label{Sec:3}
Now that we have built the master equation from the microscopic model, we can finally simulate our time-crystal by computing the time-crystal signature $\langle S_x(t)\rangle$ as a function of time, using the methods described in the previous section:
\begin{equation}
    \langle S_x(t)\rangle = \text{Tr}\left[ \rho_S(t) S_x \right].
\end{equation}

Before we begin, we would like to emphasize that the jump operators we derived differ significantly from those proposed in Ref.~\cite{lazarides2017fate}: $L_{\epsilon \epsilon'} = g(\epsilon') |V_{\epsilon\epsilon'}|\ket{\epsilon}\bra{\epsilon'}$, where $g(\epsilon') \propto \exp(-\beta \epsilon')$. Indeed, as we have already mentioned in the introduction, this approach led to an unexpected dependence of the time-crystal signature on temperature: the oscillations were damped less as the bath temperature increased. As we shall see, our model will lead to the expected dependence.

First of all we must choose the interaction operator $V$. The choice of interaction operator $V$ with the thermal bath is crucial for the interpretation of the subsequent results. We will consider two choices in this paper: $V = \sum_i \sigma_i^z$ and $V =\sum_i \sigma_i^x$, which are both relevant to model experimental platforms for the realisation of DTC. Selecting one over the other leads to different and non-trivial outcomes since both do not commute with at least one of the Hamiltonian strokes.
If the system is viewed as a collection of two-level systems in the $\sigma_z$ basis, the former choice ($z$-interaction) is effectively a dephasing channel. As we will see in the following sections, this results in dephasing during the $z$-stroke and dissipation during the $x$-stroke. The latter choice ($x$-interaction) can be seen as a dissipation channel. However, it is important to note that this channel will cause dissipation during the $\sigma_z$ stroke but only dephasing during the $\sigma_x$ stroke of the Hamiltonian since it commutes with it. From here on, we will treat these two noise channels separately, highlighting the qualitatively different outcomes that each produces.

\begin{figure}[t!]
    \includegraphics[width=16cm]{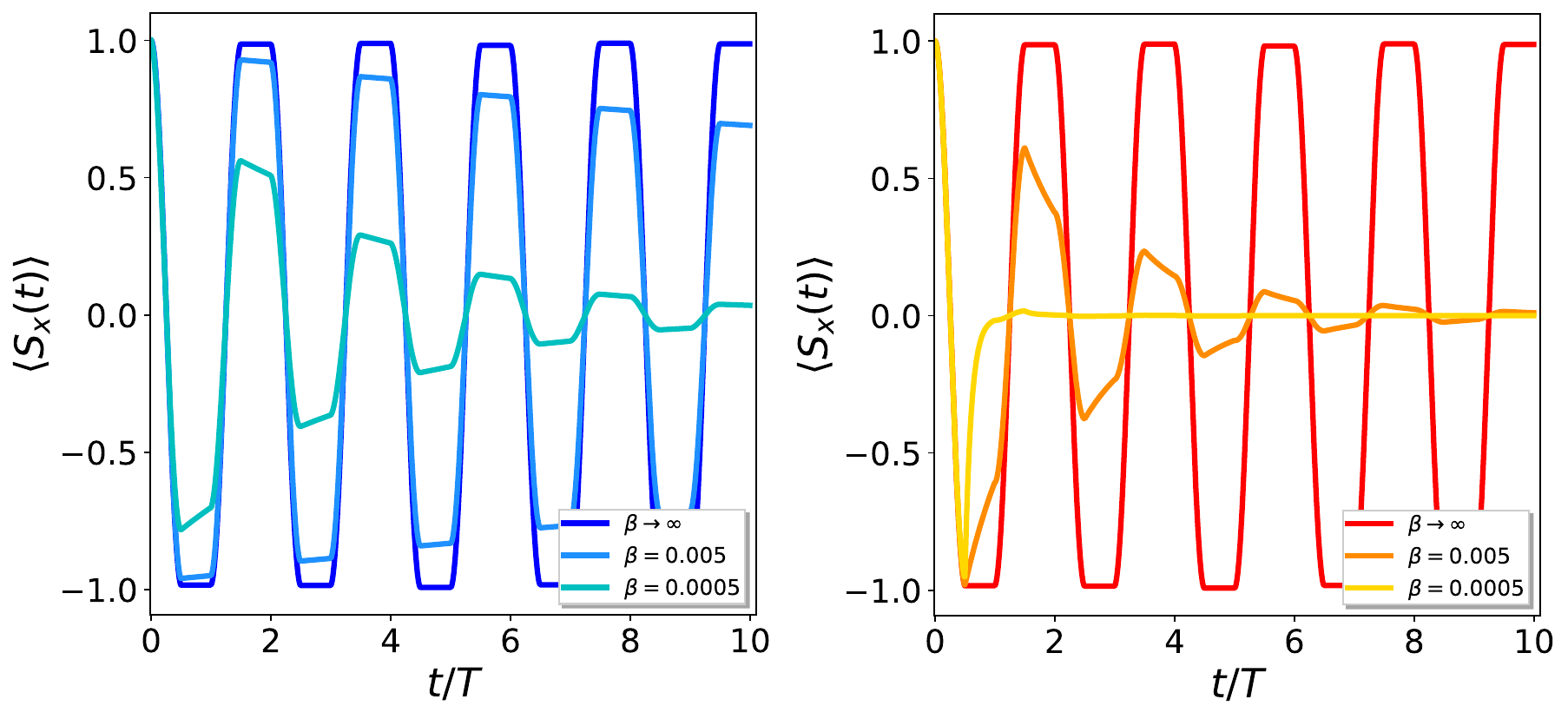}
	\caption{The time-crystal signature $\langle S_x(t) \rangle$ as a function of time for different inverse temperatures of the bath $\beta$. On the left the results in the $z$-interaction while on the right the ones for $x$-interaction.  Here the model parameters are set to $N=5$ and $\Gamma=0.01$.}
    \label{fig:timeSignature}
\end{figure}

We simulate the time-crystal made by $N=5$ sites for both interactions at different inverse temperatures $\beta$ and for $\Gamma=0.01$, obtaining the results shown in Fig. \ref{fig:timeSignature}. Our microscopic bath model leads the temperature dependence depicted in the figure, showing that the system decays faster with increasing temperature, as expected but in contrast with Ref.~\cite{lazarides2017fate}. Additionally, it seems to reveal that the oscillations remain unaffected by the thermal bath in the zero-temperature limit. However, the oscillations still decay over long timescales (for finite size crystals) or in the case of large $\Gamma$, indicating that the system is not entirely immune to dissipation.
Another notable observation from Fig. \ref{fig:timeSignature} is that, at the same temperature, the time crystal decays much faster when $V$ is an $x$-interaction compared to the case in which it is an $z$-interaction. In the latter, the effect of dephasing in the $\sigma_z$ stroke essentially leads to an imperfect spin flip. More interesting is what happens in the case of $x$-interaction, which acts as dephasing in the $x$ basis, effectively breaking the many-body localization \cite{medvedyeva2016influence}, which is the reason for the much faster decay observed.

\section{Thermodynamics}
\label{Sec:4}
\begin{figure}[t!]
    \includegraphics[width=18cm]{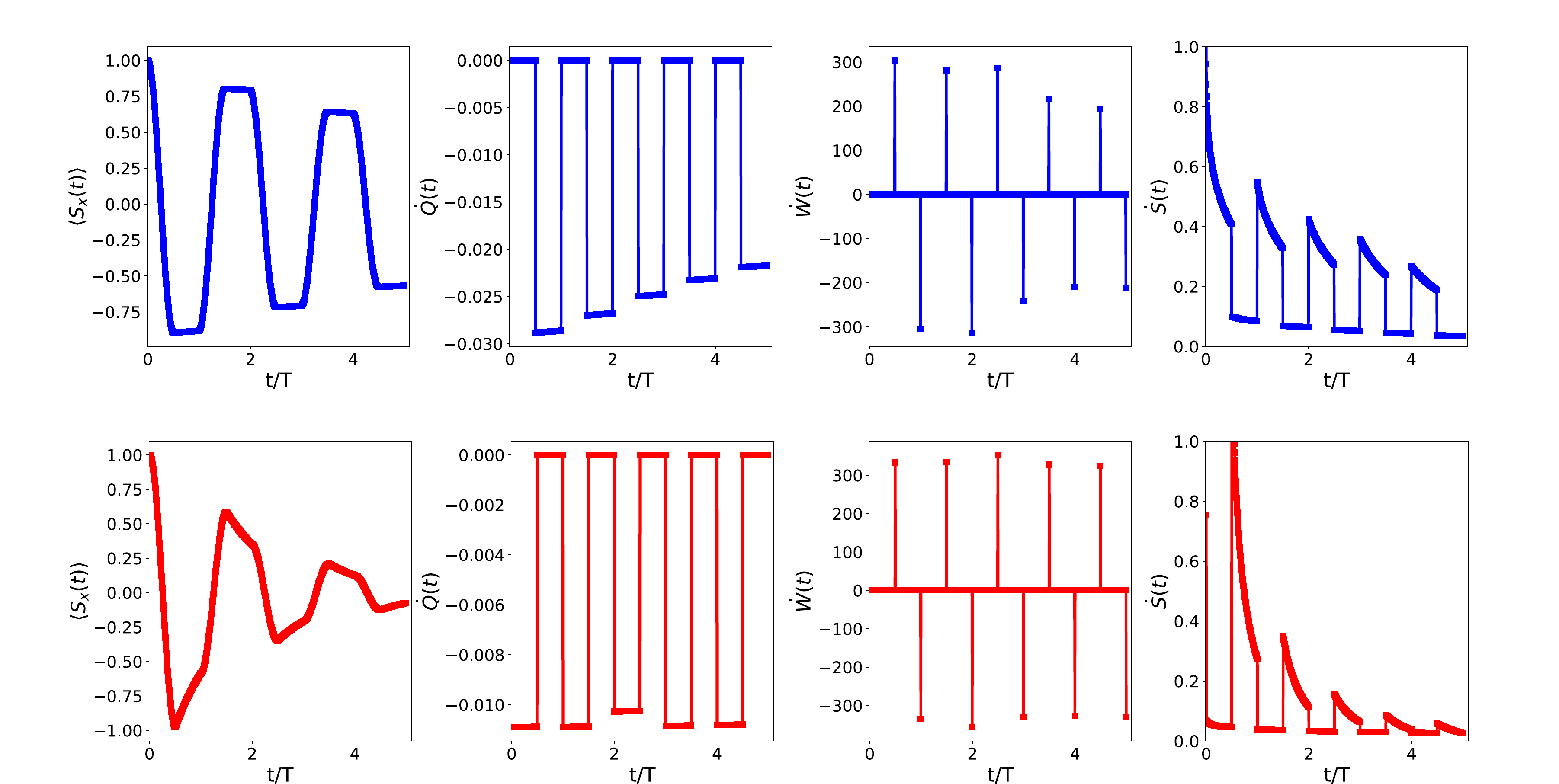}
	\caption{This figure shows the thermodynamic rates and the time-crystal signature for two different choices of interaction between the time crystal and the bath. Top: $z$-interaction, bottom:  $x$-interaction. Other parameters: $N=5$, $\beta=0.5$, $\Gamma=0.1$. As can be seen, heat dissipation occurs in two different strokes depending on the chosen interaction. Work is performed on the system when the Hamiltonians are switched, finally $\dot{S}(t)$ is always positive (with different rates depending on the stroke) in accordance with the second law.}
    \label{fig:thermo}
\end{figure}

Building upon our model of a system that exhibits a discrete time-crystal phase coupled to a thermal bath, we now turn our attention to its thermodynamic properties. With this foundation, we can explore the rates of change of key thermodynamic quantities: entropy ($\dot{S}(t)$), heat ($\dot{Q}(t)$), and work ($\dot{W}(t)$). These rates in the weak coupling limit are formally defined as follows~\cite{Alicki79}:
\begin{eqnarray}
    \dot{S}(t) &=&- \text{Tr}(\dot{\rho}\text{ log }\rho),
    \\
    \dot{Q}(t) &=& \text{Tr}(H\dot{\rho}),
    \\
    \dot{W}(t) &=&\text{Tr}(\dot{H}\rho). 
\end{eqnarray}
By these conventions, the change in the system's energy defined as $\dot{U}(t)=\frac{\text{d}}{\text{d}t} \text{Tr}(H\rho)$ fulfils the first law of thermodynamics:
\begin{equation}
    \dot{U}(t)=\dot{W}(t)+\dot{Q}(t).
\end{equation}
This formulation indicates that a positive $\dot{Q}$ corresponds to heat flowing into the system. Our analysis demonstrates that these thermodynamic properties follow the general trends depicted in Fig. \ref{fig:thermo}, for both $z$- and $x$-interactions.
Analyzing Fig. \ref{fig:thermo}, we can observe several expected behaviors. Whenever the Hamiltonian switches, work is performed on the system, resulting in the injection of heat, which is then gradually dissipated as the system cools until the subsequent switch. The entropy of the system increases monotonically, though at a decreasing rate, which can be attributed to the continuous coupling with the external environment. This increasing entropy is indicative of the system's progression towards thermal equilibrium, driven by the interaction with the thermal bath.
As observed in the case of $z$-interaction, during the $z$-stroke evolution, which corresponds to pure decoherence, we find that $\dot{Q} + \dot{W} = \dot{U} = 0$. This result is consistent with previous thermodynamic studies on pure decoherence \cite{popovic2023thermodynamics}. A similar behavior is seen for the $x$-interaction, but during the $x$-stroke. In general, the magnitude of the work features depends on the specific disorder realisation within the system.

\begin{figure}[t!]
    \includegraphics[width=18cm]{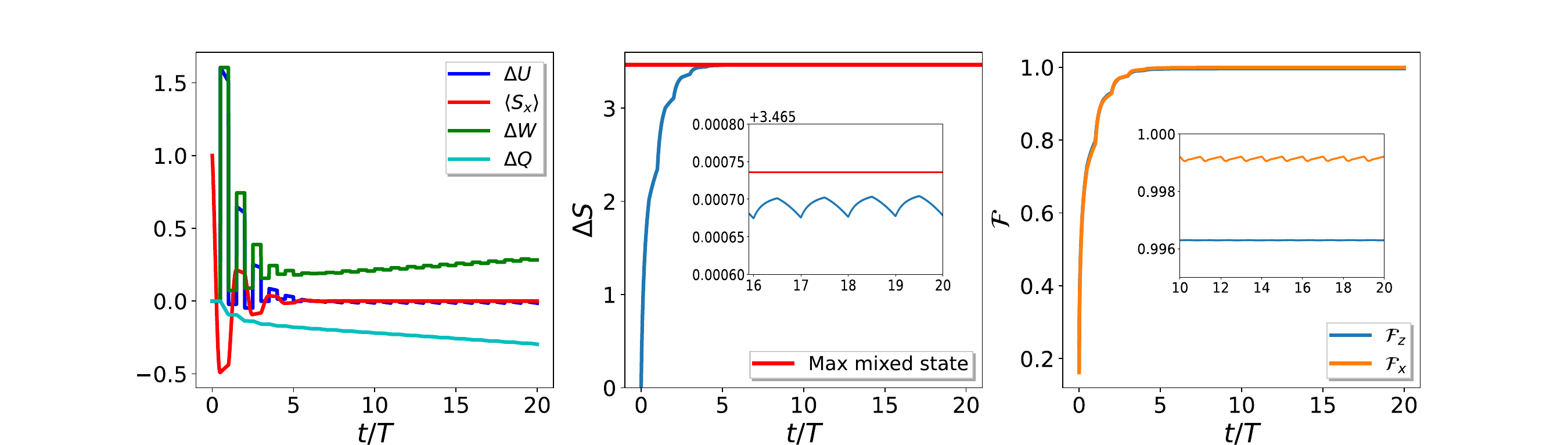}
    \includegraphics[width=18cm]{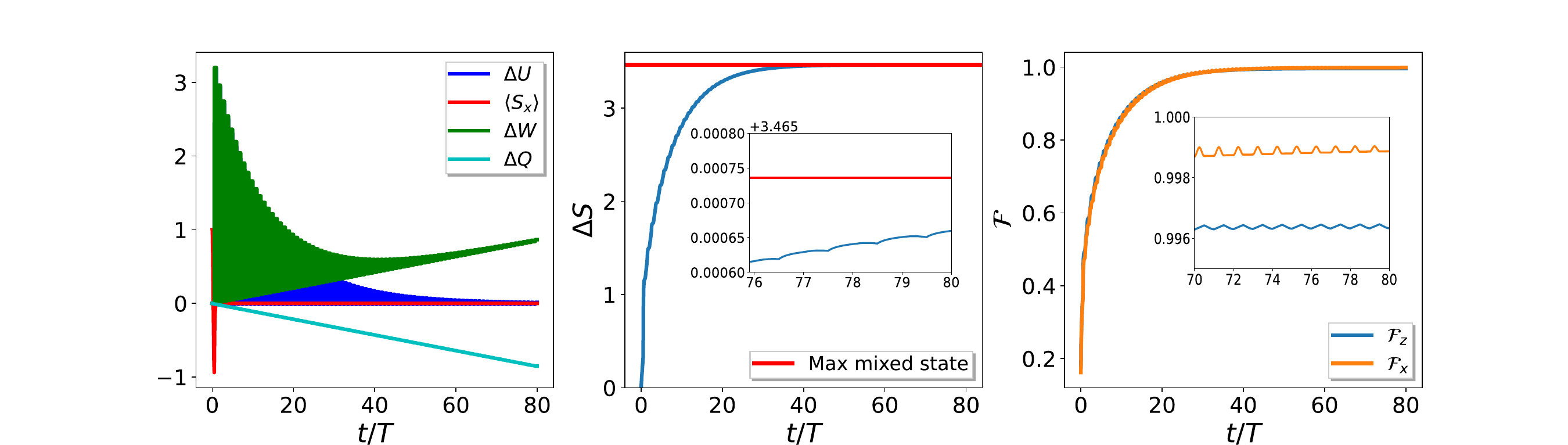}
	\caption{The death of the time crystal: Time crystal signature and thermodynamic quantities (first column), half chain Von Neumann entropy (second column) and fidelities with $H_x$ and $H_z$ thermal states (third column) as a function of time; for the DTC with: $N=5$, low $\beta$: $\beta=0.05$, $\Gamma=0.1$. Top panel: $z$-interaction, bottom panel: $x$-interaction.}
    \label{fig:death}
\end{figure}

What seems clear is that when the system is coupled to the environment, the time crystal signature inevitably decays. The unstructured environment we consider in this paper leads to the loss of crucial coherences, causing the signature of the DTC to fade away. To demonstrate this behavior we decreased the inverse temperature $\beta$ (Fig. \ref{fig:death}), and we integrated the thermodynamic quantities defined above. For high enough temperatures or coupling strength the system tends to a quasi-periodic non-equilibrium steady state.
This regime can be observed by examining the fidelity between the system state and the thermal states $ \rho = \frac{e^{-\beta H}}{\text{Tr}(e^{-\beta H})} $ (where $H$ is $H_z$ or $H_x$), and additional insights can be gained from the entropy, as shown in Fig. \ref{fig:death}.\\
Initially, the system's entropy increases during both strokes. However, in the $z$-interaction case (upper panels of Fig. \ref{fig:death}, the steady state is characterized by an increase in entropy during the $z$-stroke, followed by a corresponding decrease during the $x$-stroke. The reverse behavior is observed in the $x$-interaction case (lower panels). 
During the $H_z$ stroke, the process leads to pure dephasing, which can only increase the system's entropy. Conversely, the $H_x$ stroke is not thermalizing and can lead to a reduction in entropy, as depicted in Fig. \ref{fig:death}.


Throughout this process, when the steady state is reached, the work done on the system is always positive, while heat is consistently negative, indicating that the system acts like a dissipator: it absorbs work and releases heat into the bath. Additionally, at least in the $z$-interaction case, we find that energy flows into the system during the $H_x$ stroke and out of the system during the $H_z$ stroke while the DTC is still active. Once the DTC phase decays, energy consistently flows out of the system.
Entropy quickly reaches the value of a fully mixed state, and the system reaches a periodic non-equilibrium steady state, as evidenced by the fidelity fluctuations with the thermal states of the two Hamiltonians.
Finally, we note that in the low $\beta$ limit, thermalization occurs much more slowly in the $x$-interaction case, even though the time crystal signature decays faster, as depicted in Fig. \ref{fig:death}.

\section{Measurements}
\label{Sec:5}
In the previous sections, we explored the thermodynamic properties of a discrete time crystal. Our analysis demonstrated that, inevitably, the system decays when coupled to a thermal bath. This decay is a consequence of the environment-induced loss of coherence, which erases the signature of the DTC.

Given these findings, we now turn our attention to a critical question: can we somehow enhance the lifespan of the DTC? One promising approach involves the consideration of repeated measurement schemes. By strategically implementing these measurements, we aim to investigate whether it is possible to stabilize the DTC phase and extend its coherence time in the presence of environmental interactions. In this section, we will explore this approach, examining the potential benefits and mechanisms through which repeated measurements can influence the DTC's behavior and longevity.

The idea, which we describe in the following, is particularly suited for the selected parameters, where the time crystal oscillates between $\ket{+}$ and $\ket{-}$. Every period $T$ the system approaches one of these states and, with this in mind, we measure the total magnetization $S_x = \frac{1}{N} \sum_i \sigma_x^i$ at each period to ``reset" and purify the system. By performing these measurements, the aim is to counteract the decoherence induced by the environment and potentially stabilize the oscillatory behavior of the time crystal, thereby extending its coherence and lifespan. A similar scheme has been considered in \cite{camacho2024prolonging}, where the authors implemented a continuously monitored Floquet time crystal on a quantum computer. Their approach is based on the application of a periodical quantum-classical feedback protocol. They show that single-qubit corrections based on measurement outcomes of randomly chosen qubits are able to enhance spatiotemporal order of the DTC signal. The difference is that in our case an experimentally simpler protocol without feedback after measurement is considered, to let the system evolve on its own, without any external “help.”

Starting from a generic initial state $\rho(0)$ at time $t=0$, the state after one period of the evolution is then:
\begin{equation}
    \rho(T) = \Lambda(T) \rho(0) = \Lambda_x(T_x) \Lambda_z(T_z) \rho(0).
\end{equation}
Now, if we measure the total magnetization $S_x$, the state after the measurement is given by:
\begin{equation}
    \rho_i'(T) = \frac{\Pi_i \rho(T) \Pi_i}{p_i},
\end{equation}
where $p_i$ is the probability of measuring the eigenvalue $m_i$ of $S_x$, and $\Pi_i$ is the corresponding projector. 
The procedure is then repeated with $\rho_i'(T)$ as the initial state for the next period.
Note how the period of the measurement is the same as that of the driving and cannot hence induce a sub-harmonic response.

Furthermore, since we are interested in the time crystal signature, we need to average the observable over all possible measurement outcomes. By definition, assuming that $T < t \leq 2T$, this is given by:
\begin{align}
    \overline{\langle S_x(t)\rangle}&= \sum_i p_i \langle S_x(t)\rangle_i=\sum_i p_i \text{Tr} \biggl[S_x \Lambda(t) \rho_i'(T) \biggr]
    \equiv \text{Tr} \biggl[S_x \Lambda (t) \rho'' \biggr],
\end{align}
where we have defined $\rho''\equiv \sum_i \Pi_i \rho(T) \Pi_i$. In this way, the average is conveniently calculated by determining the statistical mean value of the state $ \Lambda (t) \rho''$, which is nothing more than the time evolution of the incoherent mixture of all possible eigenstates of the observable $S_x$. It is straightforward to see that the procedure can be iterated for $2T < t \leq 3T$ by replacing $\rho''$ with $\rho''' \equiv \sum_i \Pi_i \Lambda(T) \rho'' \Pi_i $, and so on for larger times.

\begin{figure}[t!]
    \includegraphics[width=16cm]{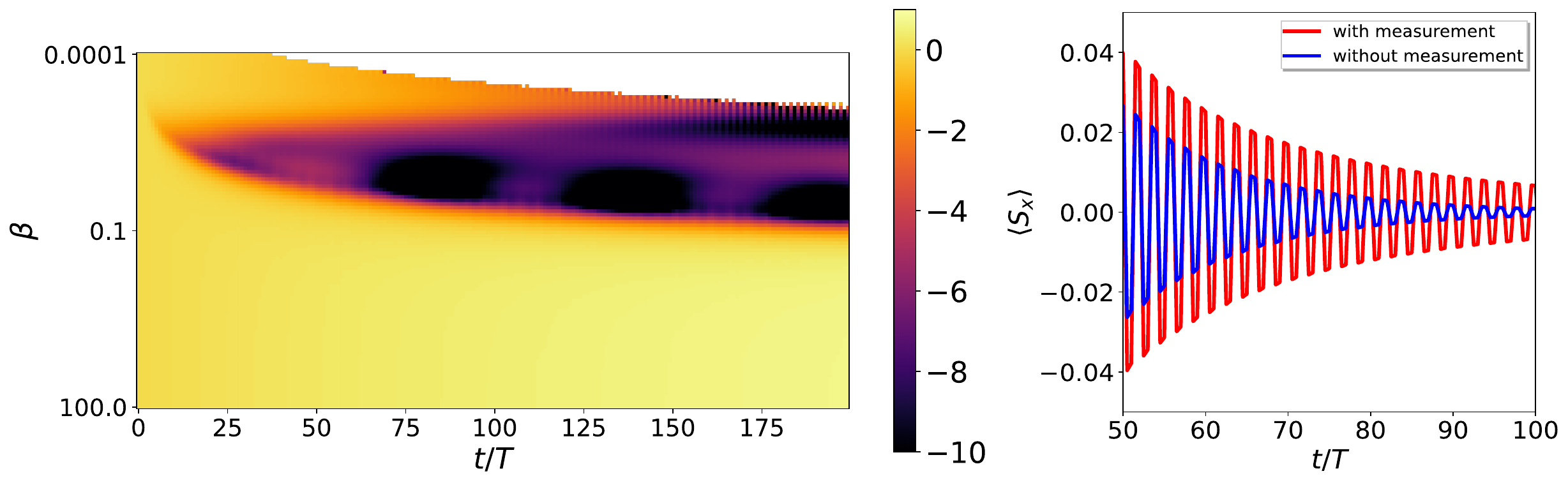}
	\caption{Left: The relative difference $E_r$ as a function of time and inverse temperature. Right: $E_r$ as a function of time for $\beta=0.05$). Other parameters: $N=5$, $\Gamma=0.01$.}
    \label{fig:relativeDiff}
\end{figure}

With this formula, we can compute the DTC signature as a function of time and inverse temperature. First, we notice that for the $z$-interaction, measurements do not improve the DTC lifespan; however, there is a stabilization of periodicity in the presence of an external field. 
More interestingly, in the case of the $x$-interaction, the results are notably different.
In Fig. \ref{fig:relativeDiff}, we have plotted the absolute relative difference between the time crystal signature without measurements and with measurements performed every $T$:
\[ E_r = \frac{ |\langle S_x \rangle_0| - |\overline{\langle S_x \rangle}|}{|\langle S_x \rangle_0| } \] 
as a function of time and inverse temperature. As can be seen, there is a moderate advantage (roughly an order of magnitude) for high temperatures and long times (darker region of the plot) whereby the mean DTC’s signature in the presence of measurements (red curve in the right panel) has a larger amplitude than the signature without measurements (blue curve in the right panel). 

This improvement can be interpreted as a sort of sub-Zeno effect caused by measurements. The quantum Zeno effect is a phenomenon in quantum mechanics in which frequent measurements can inhibit the evolution of a quantum system \cite{facchi2008zeno}. This effect is particularly significant and better understood in closed quantum systems, where continuous observation can effectively freeze the dynamics by repeatedly collapsing the wave function due to the short time quadratic behavior of the fidelity.
In open quantum systems, the situation is quite different. In general, it is not possible to determine a-priori whether a system can be frozen or not, and indeed, measurements typically have the opposite effect \cite{gough2014zeno, wu2017zeno}. However, Zeno dynamics can manifest itself as a slowed down decoherence or even stabilization of certain states due to frequent measurements. In our case, the frequency of the measurements is determined by the driving period, which is more or less of the same order as the decay time of the time crystal. This prevents a true Zeno effect from occurring, and for this reason, we refer to it as a sub-Zeno effect.

Moreover, the reason why this improvement here occurs only at high temperatures and not at low temperatures can be intuitively understood through the disordered nature of the DTC's Hamiltonians. At low temperatures, the time crystal does not evolve into the states $\ket{+}$ and $\ket{-}$ precisely. Therefore, the probability of measuring one of these magnetizations is never one, which inevitably leads to "errors" due to the measurement process. This randomness prevents a proper Zeno effect from occurring at low temperatures.

\sect{Properties of single trajectories}
As we have seen before, on average, the advantage of the measurement scheme is limited to high temperature regions. However, we can obtain interesting insights into the DTC by examining individual trajectories and in particular the outcomes of measurements. 
Using the outcomes, we define a vector $\bm{m}$, where each component represents an outcome, and its length corresponds to $\mathcal{T}$, the total number of periods driven (total time is $\mathcal{T} T$). For all intents and purposes, this vector can be considered a kind of classical spin system in the time domain, allowing us to calculate various quantities that could provide information about the time crystal.
\begin{figure}[t!]
    \includegraphics[width=14cm]{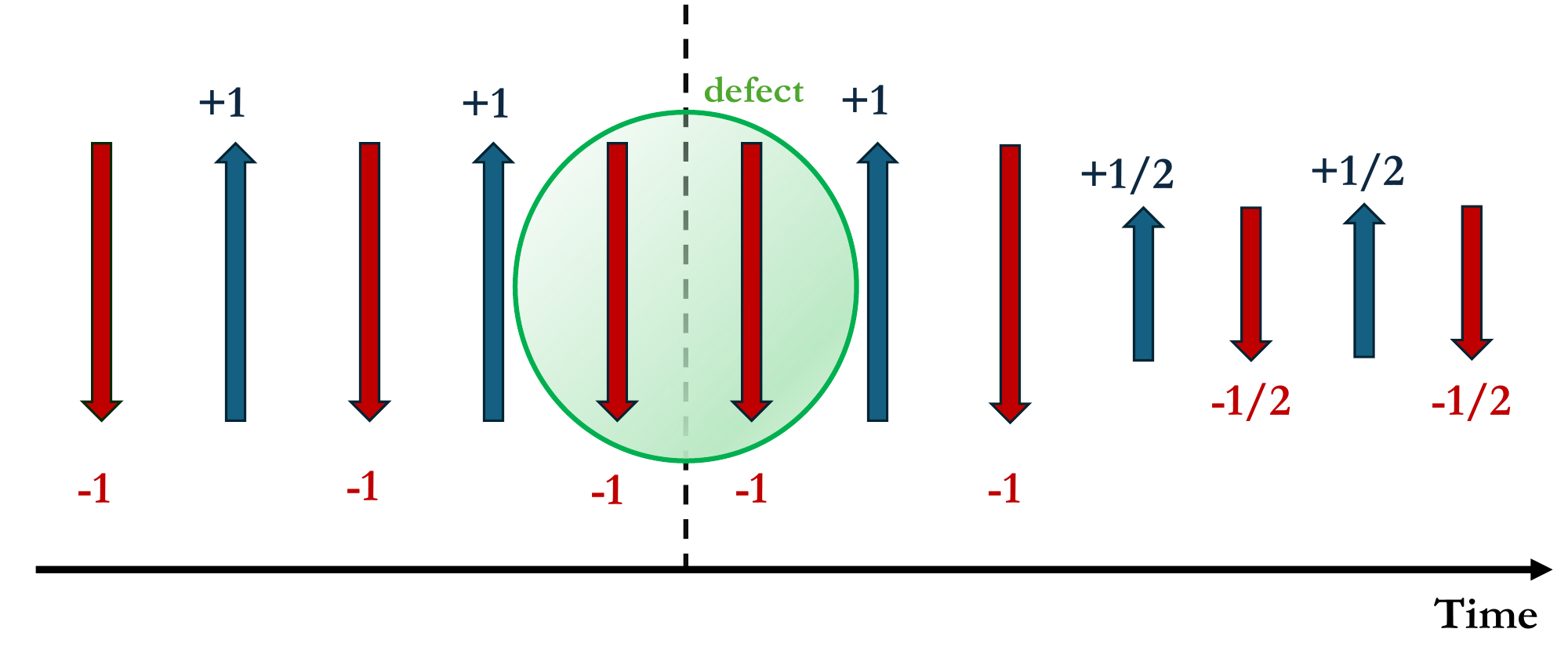}
	\caption{Pictorial example of measurement string for $N=4$. Each measurement is performed every time $T$. Negative measurement results are shown in red and positive  results in blue. When there are two consecutive measurements of the same sign we have a defect (represented in green), the dotted line divides two different domains. The possible outcomes for $N=4$ are $1,1/2,0,-1/2,1$; note that we do not consider the consecutive measurements $-1$ and $1/2$ as defects.}
    \label{fig:meas}
\end{figure}
In principle, in the zero-temperature limit for the chosen parameters, as we measure $S_x$ each period, we expect to observe an alternation of 1 and -1, except for possible errors due to the imperfect spin flip pulses. This approach enables us to gain a deeper understanding of the DTC's behavior under different conditions. We can define a time-staggered magnetization:
\begin{equation}
    M\equiv \frac{1}{\mathcal{T}}\sum_{i=1}^\mathcal{T}m_i (-1)^i,
\end{equation}
which would take the maximum values $\pm 1$ in the case of perfect alternation of signs. In general, it assumes values between -1 and 1: due to the coupling with the bath, errors may occur, manifesting as intermediate measurements (e.g., in the case of four spins, we may measure 1/2 instead of 1) or, in the most extreme cases, as changes in sign.
We refer to these changes in sign as defects: a defect occurs whenever the magnetization at two consecutive times ($m_i$, $m_{i+1}$) does not change sign. Thus, we can define the defect density as:
\begin{equation}
    dw=\frac{1}{\mathcal{T}}\sum_{i=1}^\mathcal{T}  \frac{1+\text{sign}(m_i)\text{sign}(m_{i+1})}{2}.
\end{equation}
This quantity can take values between 0 and 1. In the case of perfect alternation of signs, it is 0, while in the case of completely aligned signs, it reaches 1.
Furthermore, we can define domains as sets of consecutive defect-free measurements. We therefore define the mean size of these domains as
\begin{equation}
    A=\frac{1}{\#\text{domains}}\sum_{\text{domains}} l(\text{domain}),
\end{equation}
where $l$ indicates the length of a domain. This quantity is related to the width of the measurement string auto-correlation function, since the latter measures the characteristic time scale over which the signal remains correlated with itself, indicating how quickly the correlations in the signal decay.

\begin{figure}[t!]
    \includegraphics[width=16cm]{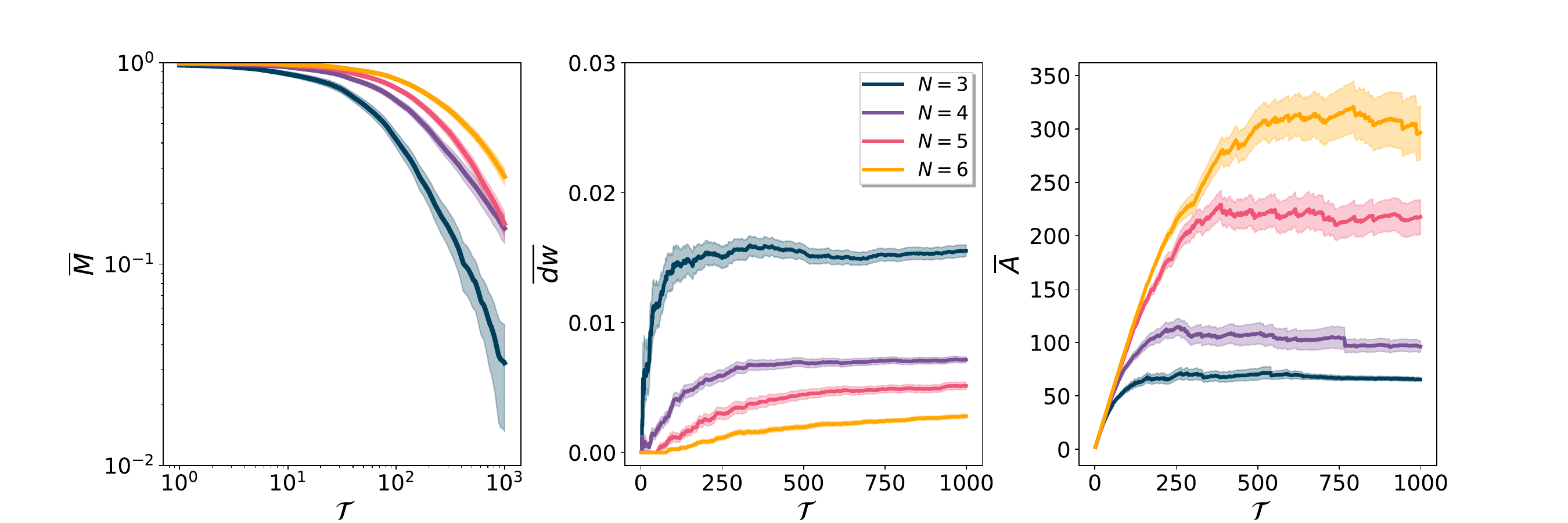}
	\caption{Time-staggered magnetization of the time crystal, defect density and average domain size as function of $\mathcal{T}$ for different time crystal sizes for $\beta=10$ and $\Gamma=0.01$. 
}
    \label{fig:IterativeMean}
\end{figure}

In Fig.~\ref{fig:IterativeMean}, the quantities defined earlier are shown as functions of the measurement string length $\mathcal{T}$, averaged over 100 different realizations and for various system sizes. As can be observed, for the chosen temperature, the staggered magnetization $\overline{M}$ (where the bar indicates the average over the different realizations) decays for each system size following what appears to be a power law. More interestingly, both $\overline{dw}$ and $\overline{A}$ seem to settle on distinct plateaus. In the case of $dw$, the height of the plateau decreases with increasing system size, while for $A$, it appears to increase.
The measurement string thus seems to form distinct domains that indicate the underlying time crystal phase, suggesting an increasing robustness of the time crystal with system size. 

\begin{figure}[t!]
    \includegraphics[width=16cm]{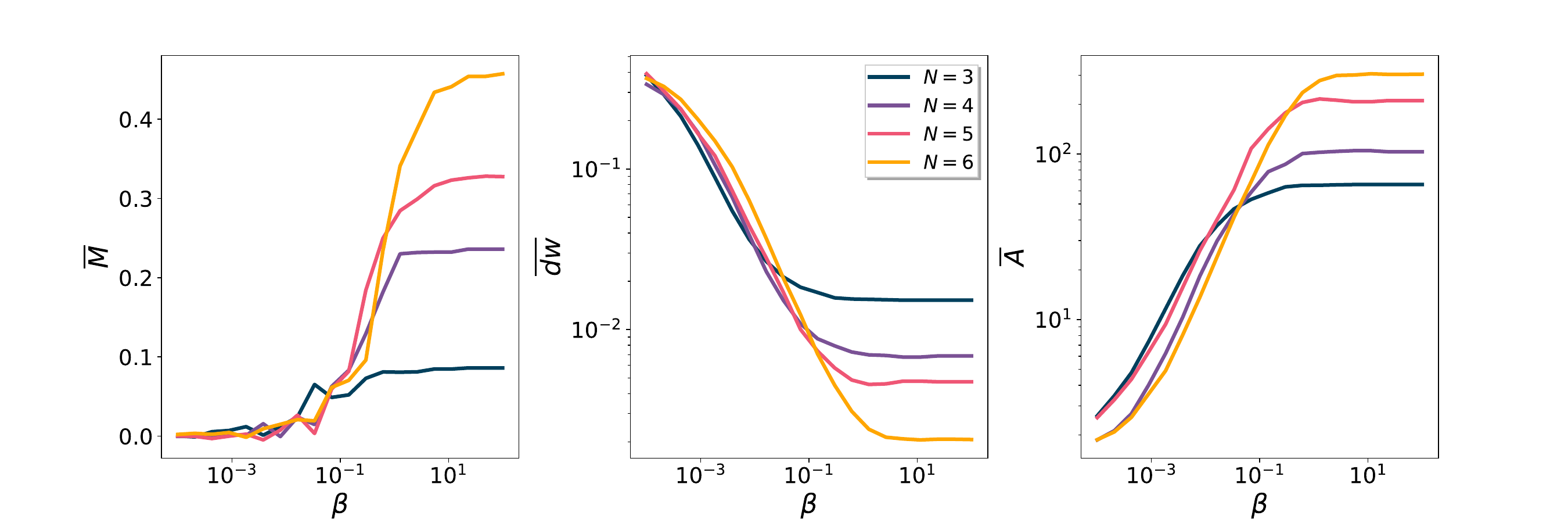}
	\caption{Staggered magnetization, defect density and average domain size as a function of $\beta$ for different time crystal sizes for $\mathcal{T}=500$ and $\Gamma=0.01$.}
    \label{fig:FOMvsBeta}
\end{figure}

A natural question that arises at this point is what happens when the temperature of the system is varied.
To address this question, we varied the inverse temperature $\beta$ while keeping $\mathcal{T}=500$ fixed. As shown in Fig.~\ref{fig:FOMvsBeta}, the system exhibits interesting behavior. Two distinct regimes can be observed: a disordered one, which occurs for small $\beta$ values and characterized by a non-zero defect density and almost unit-sized domains dimension, and an ordered regime, where the defect density tends to zero and the domains dimension becomes macroscopic. 
Interestingly, information about the time crystal phase can be extracted directly from the results of repeated measurements, without the need for reconstructing averages, as even a few trajectories already provide the relevant information. This feature could be particularly advantageous in experimental applications.

\section{Conclusions}
\label{Sec:6}
In this work, we examined the properties of a discrete-time crystal (DTC) coupled to a global bosonic thermal bath, in the weak-coupling and Markovian limit. In particular, we have derived the proper form of the jump operators from the microscopic model and we have calculated some relevant thermodynamic quantities, such as heat, work and entropy.

Our analysis demonstrates that, for any temperature of the thermal bath, the coupling to the latter suppresses the subharmonic response characteristic of DTC. Therefore, we have focused on trying to stabilize these systems, specifically using a simple protocol involving repeated projective measurements. As expected, on average, these measurements have a moderate or even a negative effect, since they essentially mimic decoherence. However, when considering individual trajectories, we observed a stabilization of the DTC dynamics, characterized by time intervals of stability defined by certain parameters associated with the measurement outcomes. We thus observe that the signature of the DTC remains stable over long times.
Notably, the size of these stability intervals appears to increase as the system size grows, suggesting that the protocol may offer some robustness in the thermodynamic limit. Nevertheless, further analysis with larger sizes would be needed to determine whether there is a phase transition or simply a crossover between a stable and a suppressed DTC.
These findings may be valuable for future experimental implementations, where decoherence due to the environment plays a key role in the dynamics and stability of these systems.
\section{Data/code availability}
The codes used to generate the data for this article and the data themselves are available from the corresponding author upon reasonable request.
\bibliography{biblio.bib}{}
\begin{acknowledgments}
G.C. acknowledges hospitality from Queen's University Belfast, G.C. and G.B. acknowledge support from the PRIN 2022 - 2022XK5CPX (PE3) project funded within the program ‘PNRR Missione 4 - Componente 2 - Investimento 1.1 Fondo per il Programma Nazionale di Ricerca e Progetti di Rilevante Interesse Nazionale (PRIN)'
\end{acknowledgments}



\clearpage

\end{document}